\documentclass[reprint,aps,prl, 
balancelastpage,a4paper,floatfix,superscriptaddress]{revtex4-2}

\usepackage[margin=0.75in]{geometry}
\usepackage[T1]{fontenc}
\usepackage{graphicx}
\usepackage{float}
\usepackage{amsmath,amssymb,bm}
\usepackage{xcolor}
\usepackage{booktabs}
\usepackage{enumitem}
\usepackage{dcolumn}
\usepackage{url}
\usepackage{hyperref}
\usepackage{hycolor}
\hypersetup{colorlinks=true,citecolor=[rgb]{0,0.18,0.68}, 
urlcolor=[rgb]{0,0.18,0.68}, linkcolor=[rgb]{0,0.18,0.68}, final = true}
\usepackage[rightcaption]{sidecap}
\usepackage{physics}
\usepackage{mathtools}
\usepackage{relsize}
\usepackage{mathptmx}
\usepackage{cfr-lm} 
\usepackage{lmodern}

\usepackage{siunitx}
\DeclareSIUnit\angstrom{\text {Å}}

  \def\be{\begin{equation*}}
  \def\ee{\end{equation*}}
  \def\ba{\begin{eqnarray}}
  \def\ea{\end{eqnarray}}
  
  \def\eref#1{Eq.(\ref{#1})}
  \def\fref#1{Fig.~\ref{#1}}

  \def\bt{\textrm} 

  \def\nsi#1{\noindent\textit{\bt{#1.---}}}
  \def\nsq#1{\noindent\textit{\bt{#1---}}}
  \definecolor{or}{RGB}{234,142,53}
  \definecolor{gr}{RGB}{150,150,150}
  \definecolor{bl}{RGB}{54,152,187}

  \def\dG{\Delta G}
  \def\ddG{\Delta \Delta G}
  
  \def\rb{\mathbf{r}}
  
  \def\dN{S}
  \def\dNi{S_{i}}
  \def\dNm{S_{m}}
  \def\dNn{S_{\rm n}}

  \def\CA{\mathrm{C}_{\alpha}}

  \newcommand{\ie}{\textit{i.e.}}

  \definecolor{YKB}{rgb}{0.00,0.18,0.65}

\begin{document}

\title{
AI-predicted protein deformation encodes energy landscape perturbation}

  \author{John M. McBride}
    \email{jmmcbride@protonmail.com}
    \affiliation{Center for Soft and Living Matter, Institute for Basic Science, Ulsan 44919, South Korea}
  \author{Tsvi Tlusty}
    \email{tsvitlusty@gmail.com}
    \affiliation{Center for Soft and Living Matter, Institute for Basic Science, Ulsan 44919, South Korea}
    \affiliation{Departments of Physics and Chemistry, Ulsan National Institute of Science and Technology, Ulsan 44919, South Korea}

\begin{abstract}
  AI algorithms have proven to be excellent predictors of protein structure, but whether and how much these algorithms can capture the underlying physics remains an open question.
  Here, we aim to test this question using the Alphafold2 (AF) algorithm: 
  We use AF to predict the subtle structural deformation induced by single mutations, quantified by strain, and compare with experimental datasets of corresponding perturbations in folding free energy $\ddG$.
  Unexpectedly, we find that physical strain alone -- without any additional data or computation -- correlates almost as well with $\ddG$ as state-of-the-art energy-based and machine-learning predictors. 
  This indicates that the AF-predicted structures alone encode fine details about the energy landscape. In particular, the structures encode significant information on stability, enough to estimate (de-)stabilizing effects of mutations, thus paving the way for the development of novel, structure-based stability predictors for protein design and evolution.
\end{abstract}

\maketitle

  AI has ushered in a revolution in structural biology, yet we are still in uncharted waters~\cite{thonm21,subnm22}. In particular, it is not clear whether AI algorithms that predict protein structure from sequence, such as AlphaFold (AF)~\cite{jumna21} or RoseTTAFold~\cite{baesc21}, owe their unprecedented accuracy to highly sophisticated pattern recognition or these algorithms can capture some of the many-body physics underlying protein folding. Recent studies provide evidence suggesting that AF has learned an \emph{effective} energy functional that is searched in order to accurately predict the native structure~\cite{ronpr22}, even if it includes uncommon structural motifs~\cite{herpn23}.
  
  A stringent test for whether an AI algorithm has learned the actual \emph{physical} energy landscape would be the capacity to probe, from the predicted structure alone, changes in the thermodynamic free energy ($\ddG$) due to single mutations. This would indicate that the predicted structure encodes fine details about the physical energy landscape. Besides this fundamental interest, such capacity may be impactful in applications: Of particular importance for protein design and sequence generation~\cite{golar18,kuhnr19,sgael23,malcur2023}, and for protein evolution~\cite{tokco09,otabi18,maupr23} is the ability to predict whether a given protein sequence will lead to a stably-folded structure~\cite{jumna21,baesc21}. 
  AF has been reported to predict folded structures for proteins that are not stable~\cite{mcbpr23,buens22,pakpo23}. Despite this, some analyses suggest that AF and RoseTTAFold can be used for predicting stability changes upon mutation~\cite{mcbpr23,akdns22,manbi22,iqbjc22,pecjc23}.
  All this motivates us to systematically investigate here the question of whether AI-predicted structures can be used to infer changes to free energy landscapes.

  To this end, we study a curated subset of \num{2499} measurements of stability change ($\ddG$) due to single mutations, taken from ThermoMutDB (TMDB)~\cite{xavna21}.
  Strain is a simple general measure of deformation in proteins~\cite{mcbjs24,eckrmp19,mitpn16,tluprx17,dutpna18,mcbmb22}.
  We find that the deformation upon mutation -- as measured by the effective strain~\cite{mcbpr23} -- correlates well with $\ddG$. We show it is essential to average over an ensemble of multiple AF-predicted structures to get precise estimates of strain due to mutation, and that most of the relevant information is gleaned from the residues within \SI{15}{\angstrom} of the mutated residue.
  Our initial motivation was to examine whether the AF structures encode any information about stability. Surprisingly, we found that correlations between strain and $\ddG$ compare well against those obtained using state-of-the-art $\ddG$ predictors, suggesting that AF predictions are highly informative of stability changes. We propose that new energy-based force fields can be developed that may provide a mechanistic understanding of the effects of mutations on stability. Since stability measurements are easier than structure determination, such a development could settle the question of whether AI algorithms have truly learned the physics of protein folding by testing on massive stability datasets~\cite{ronpr22,herpn23,tsuna23}.

\begin{figure*}
\centering
\includegraphics[width=0.95\textwidth]{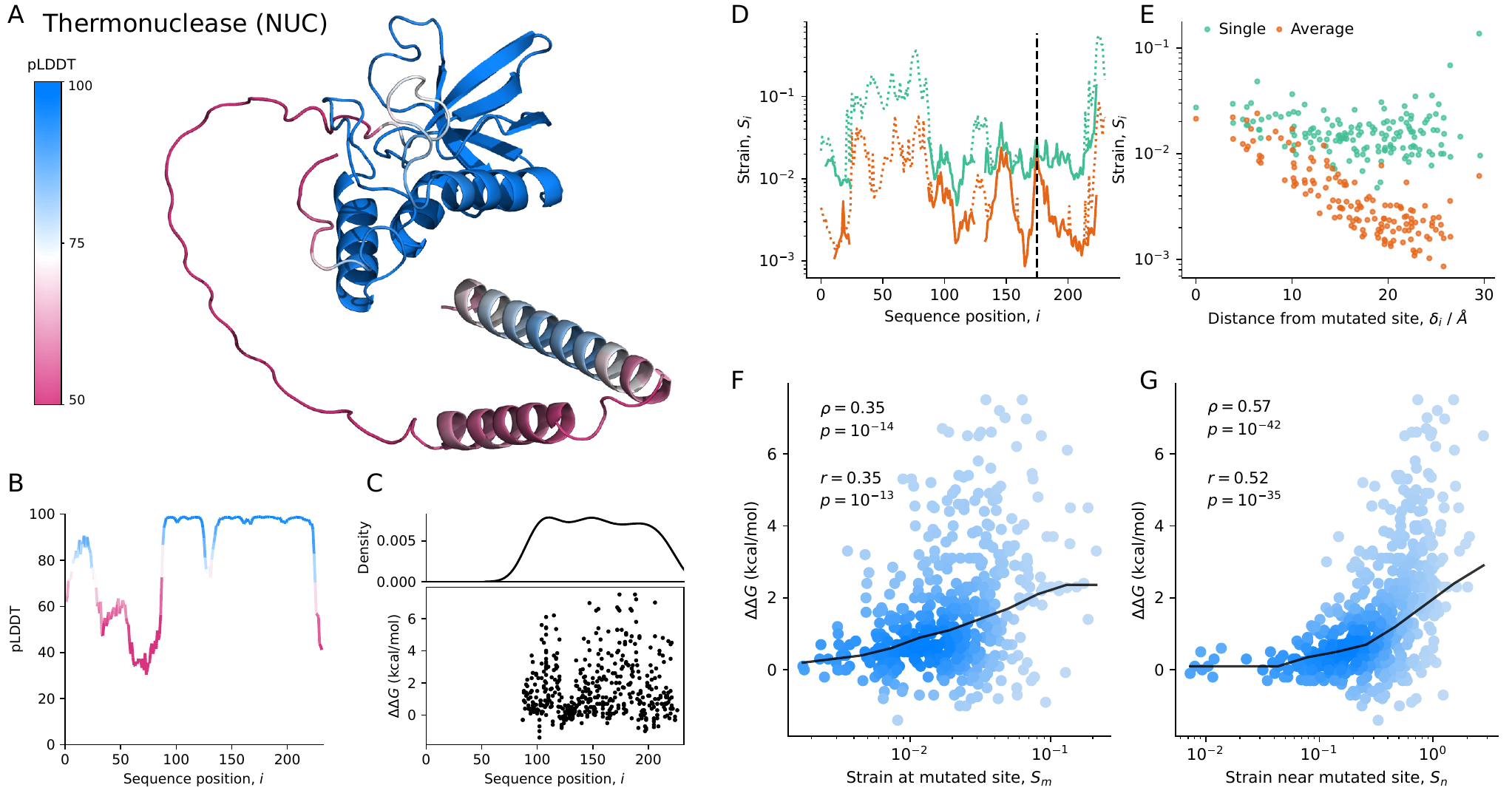}
\caption{\label{fig:fig1}
  \textbf{Strain calculated using AF-predicted structures correlates with $\ddG$.}
  A: AF-predicted structure of thermonuclease from \textit{staphylococcus aureus} (NUC) -- the most common protein in TMDB. Residues are colored according to pLDDT (AF-predicted confidence score). 
  B: pLDDT per residue.
  C: Distribution of \num{491} mutation sites along the sequence, and corresponding changes in stability $\ddG$.
  D: Strain upon mutation (A176G, $\ddG = \SI{2.4}{kcal/mol}$) per residue; the mutated site is indicated by the black dashed line. The strain calculation either includes (dotted line) or excludes (solid line) residues with pLDDT $< 70$, and uses either a single pair of structures (green) or pairs of averaged structures (orange).
  E: Strain as a function of distance from the mutated site, $\delta_i$, for single pairs and ensemble-averaged pairs of structures.
  F-G: Empirical $\ddG$ vs. strain at the mutated site, $\dNm$ (F), and the sum of strain over all residues within \SI{15}{\angstrom} of the mutated site, $\dNn$ (G). Solid line shows the median; Spearman's $\rho$, Pearson's $r$, and corresponding $p$ values are shown; circles are shaded by density. 
}
\end{figure*}


\begin{figure*}
\centering
\includegraphics[width=0.95\textwidth]{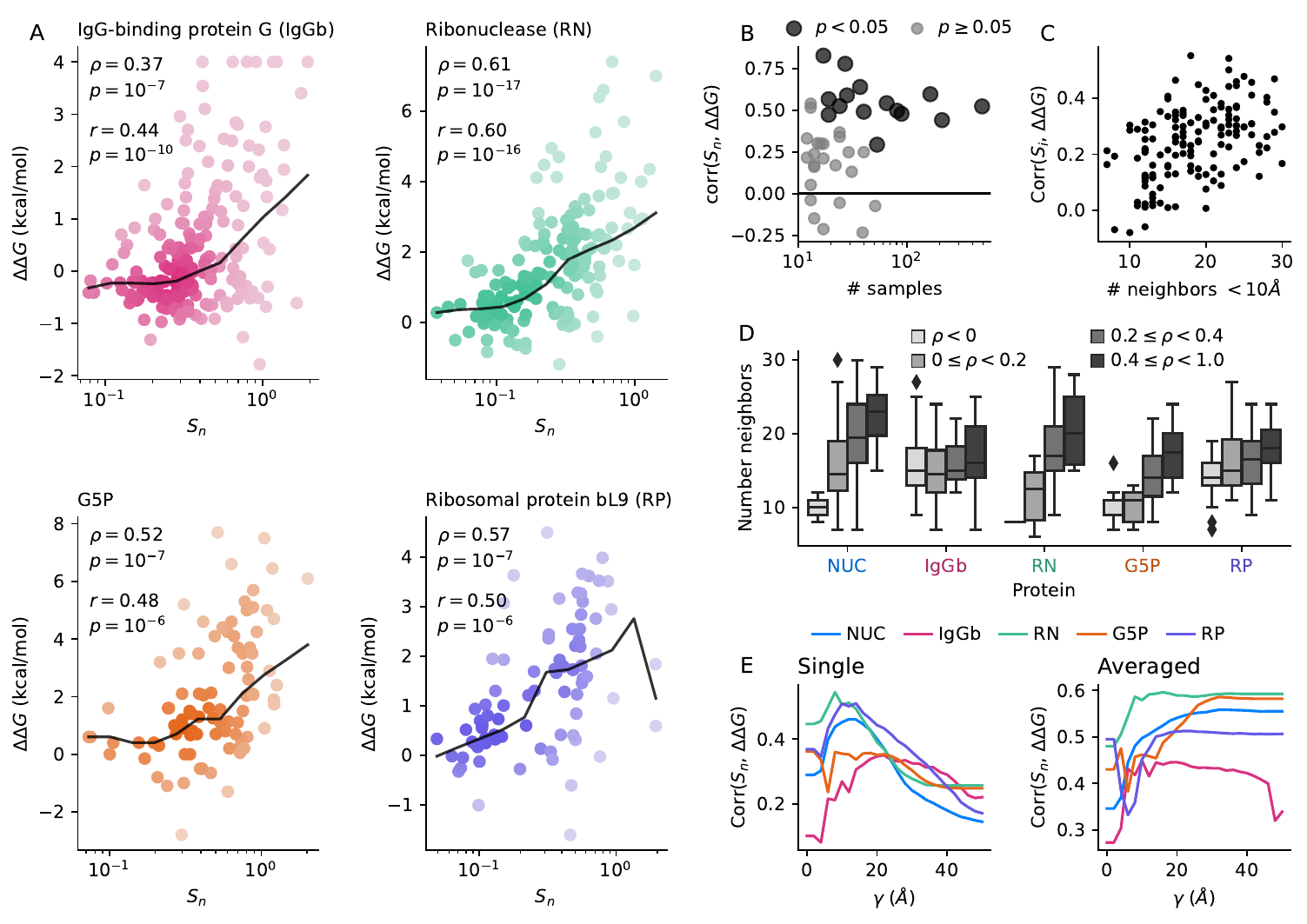}
\caption{\label{fig:fig2}
  \textbf{When and why does $\dNn$ correlate with $\ddG$?}
  A: Change in stability $\ddG$ against strain near mutated residue $\dNn$ for four proteins (the second to fifth most common in TMDB; the most common one, NUC, is shown in \fref{fig:fig1}); Spearman's $\rho$, Pearson's $r$, and corresponding $p$ values are shown; black line is the median; circles are shaded by density.
  B: $\dNn{-}\ddG$ correlation for each of the \num{40} most common proteins in TMDB as a function of the number of samples; results with $p<0.05$ are shown by dark circles.
  C: $\dNi{-}\ddG$ (Pearson's) correlation for each residue $i$ as a function of the number of neighbors (for NUC only).
  D: Distributions of numbers of neighbors, grouped by $\dNi{-}\ddG$ Spearman's correlation, for the five most common proteins.
  E: $\dNn{-}\ddG$ (Pearson's) correlation for each residue $i$ as a function of the neighborhood threshold value $\gamma$ for calculating $\dNn$, for both pairs of single structures and averaged structures, for five proteins.
}
\end{figure*}

\nsi{Strain correlates with $\ddG$ in Thermonuclease}
  We first examine thermonuclease (NUC, Uniprot ID, \texttt{P00644}; \textit{staphylococbus aureus}) -- the protein that has the highest number of mutants in TMDB (\num{491} after applying controls, Appendix A).
  NUC consists of a folded region (starting around K88) and an extended disordered region near the N-terminus (\fref{fig:fig1}A), indicated by the low pLDDT (AF-predicted confidence score) values (\fref{fig:fig1}A-B). The NUC mutants are all sampled from the folded domain (\fref{fig:fig1}C). Note that we define $\ddG$ such that an increase in $\dG$ upon mutation is \textit{destabilizing}.

  To measure deformation, we calculate effective strain (ES; Appendix B) per residue, $\dNi$, between wild-type (WT) and mutant structures predicted by AF (\fref{fig:fig1}D).
  Disordered residues always show high ES (\fref{fig:fig1}D, dotted lines) due to prediction noise, regardless of mutations~\cite{mcbpr23}. We therefore exclude residues whose pLDDT $< 70$ from the ES calculations (\fref{fig:fig1}D, solid lines). 
  Even without disordered residues, if we only compare two static structures (\fref{fig:fig1}D-E, teal), we still see residual ES in regions far from the mutated site. This occurs since regions in proteins with high flexibility tend to have high variability across repeat AF predictions (Supplementary Material [SM] Fig. 1)\cite{supp}. To achieve a more accurate estimate of deformation \textit{due to mutation}, we calculate deformation using `averaged' AF structures (Appendix B) ~\cite{mcbpr23}. 
  This drastically reduces ES in most regions (which originates chiefly from noise and fluctuations), except for regions near the mutated site (\fref{fig:fig1}D-E, orange). 
  
  Using this more precise measure of deformation due to mutation, we find a significant correlation (Spearman's $\rho = 0.35$) between strain at the mutated site $\dNm$ and change in stability $\ddG$ (\fref{fig:fig1}F).
  This correlation is even higher (Spearman's $\rho = 0.57$) when calculating the sum of strain over all residues within a spherical neighborhood of radius $\gamma = \SI{15}{\angstrom}$ around the mutated site, $\dNn$ (\fref{fig:fig1}G).
  See SM~Fig.~2 for similar analyses without excluding low pLDDT residues and without using average structures. Our rationale for choosing $\gamma = \SI{15}{\angstrom}$ will become clear in a following section. For now, we highlight that, in this particular example (NUC) it appears that AF-predicted deformation correlates quite well with empirical measurements of changes in stability.\\

\nsi{Strain correlates with $\ddG$ within protein families}
  We expand our analysis of NUC to more protein families, again focusing on the families that have the highest coverage in TMDB. 
  For the second-to-fifth most common proteins, we find correlations between stability change and local deformation ranging from $0.39 \leq \rho \leq 0.61$ (\fref{fig:fig2}A). 
  Extending this analysis to the 40 most common proteins reveals that there are insufficient samples to show significant strain-stability correlations in most cases.
  Nevertheless, in the 16 correlations that are statistically significant, the average Spearman's $\rho$ is $0.54$ (\fref{fig:fig2}B) with an overall range \numrange{0.29}{0.78}.
  We see that for all cases with sufficient data, there appears to be a consistent correlation between strain and changes in stability.\\

\nsi{Determinants of strain-stability correlation}
  To better understand why strain is correlated with changes in stability, we examine the correlations between strain at individual residues $\dNi$ (not necessarily the mutated ones) and $\ddG$, and compare this with the number of neighbors within \SI{10}{\angstrom} of each residue $i$. For NUC, we find that mutations in buried regions (those with many neighbors) tend to have an outsize impact on stability (\fref{fig:fig2}C), as expected, given the standard paradigm of buried residues having low mutation rates~\cite{echnr16}.
  In general one expects mutations of buried residues to affect more bonds and therefore inflict larger stability changes.
  Indeed, across the five most common proteins in the TMDB, we see a clear trend whereby the residues with the highest strain-stability correlations are amongst the most buried within that protein (\fref{fig:fig2}D).

  Evidently, when only a few points are sampled, the resulting correlation is not particularly informative. Also, in a large protein, more points are needed to achieve sufficient sampling. 
  This is demonstrated most clearly in IgG-binding protein G (IgGb, \texttt{P06654}): this protein has a length $L=448$, so even though we have \num{225} samples, many buried residues do not correlate with stability changes  (\fref{fig:fig2}D). This is because
  many regions have no mutations sampled from them, so deformation remains low no matter how buried the residues are. 
  Another complication arises from the abundance of disordered regions (SM Fig.~3) in which AF appears to have little
  capacity to predict mutation effects, while well-folded regions are small. As a result, for IgGB the link between number of neighbors and effect on stability is weak.
  This case highlights the need for a nuanced approach to understanding the relationship between strain and changes in stability.\\

  \nsi{Range of informative residues}
  We find that the $\dNn{-}\ddG$ correlation increases with $\gamma$, the radius of neighborhood used to calculate $\dNn$, up to about $10 \leq \gamma \leq \SI{20}{\angstrom}$, depending on the protein family (\fref{fig:fig2}E).
  This optimal length scale of \SI{\sim 15}{\angstrom}, could be the outcome of two possible effects:
  
  First, we note that the residues that are most informative of stability changes are the most buried ones (\fref{fig:fig2}C), and not necessarily the mutated residues. Hence, the optimal $\gamma$ needs to be large enough to include some of these informative, buried residues in the $\dNn$ calculation. In TMDB, for example (SM~Fig.~4), the average distance between buried and mutated residues is \SI{10}{\angstrom}. This length scale can be rationalized by a simple geometric argument: the average distance to the center of a sphere (the most buried part) of diameter $d$ is $r_0 = (3/8)d$;
  the average volume of an amino acid is \SI{\sim 144}{\angstrom^{3}}~\cite{perej86}, so in a folded, approximately spherical domain of \num{\sim 100} amino acids, $r_0 \sim \SI{11.3}{\angstrom}$.

  Second, when we compare pairs of single AF-predicted structures, which exhibit significant noise unrelated to mutations (SM~Fig.~1), strain-stability correlations reach a maximum as a function of $\gamma$ and then decrease; this decrease is not observed for averaged structures (\fref{fig:fig2}E).
  This is consistent with reports that AF strain predictions are usually indistinguishable from noise after about \SI{15}{\angstrom}, while this range could be extended to \SI{25}{\angstrom} (and noise reduced by a factor of 4) by averaging over structures~\cite{mcbpr23,mcbjs24}.

  In summary, we propose that the reason for the maximum information from strain around a range of \SI{15}{\angstrom} is due to both the need to increase $\gamma$ to include buried residues and also because long-range mutation effects may be masked by prediction noise. 
  In this context, we note that conformational dynamics of proteins were reported to exhibit correlations with similar and even longer ranges, as demonstrated by normal mode analysis~\cite{tanpl20,tanprl21,mcbjs24}. However, one needs to remember that displacement and strain -- the tensorial derivative of the displacement -- may show different correlation lengths~\cite{mcbjs24}.

\nsi{Strain correlates with $\ddG$ across protein families}
  We find a moderate correlation ($\rho = 0.36$) between $\dNn$ and $\ddG$ when comparing all the measurements together.
  This correlation is expected to be limited due to the inability of strain to differentiate between stabilizing and destabilizing mutations. Since \eref{eq:1} measures the absolute relative change in distances, $\dN \geq 0$ by definition, and thus the strain is invariant to reversing the reference and target structures.
  Indeed, we find a somewhat higher correlation ($\rho = 0.41$) between $\dNn$ and the magnitude of stability changes, $|\ddG|$.
  This indicates that more generally, large structural changes lead to large stability changes, independent of the sign of the change.

\begin{figure}
\centering
\includegraphics[width=0.95\linewidth]{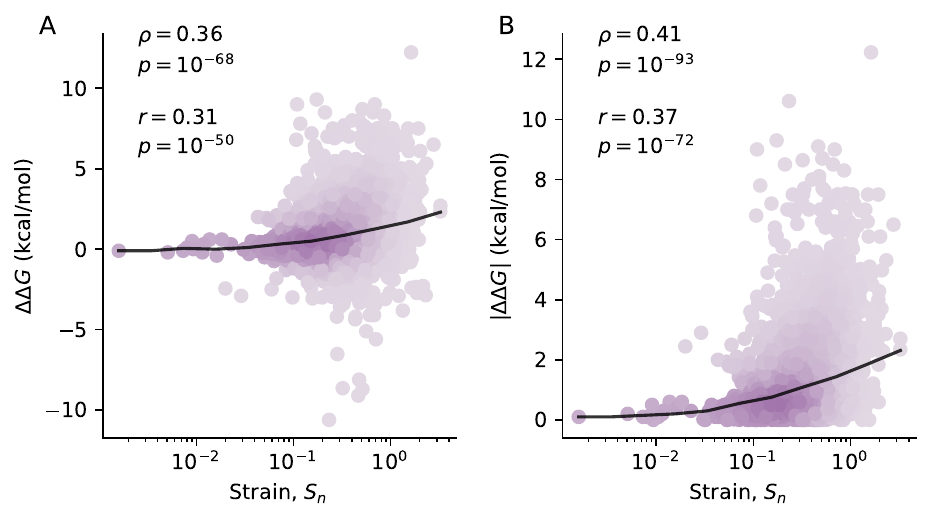}
\caption{\label{fig:fig3}
  \textbf{$\dNn{-}\ddG$ correlations across proteins.}
  Strain near mutated site $\dNn$ vs. $\ddG$ (A) and magnitude of stability change $|\ddG|$ (B) for all mutations in our reduced TMDB sample of \num{2499} unique mutants. , Pearson's $r$, and corresponding $p$ values are shown.
}
\end{figure}

  We do not expect a simple mapping between $\dNn$ and $\ddG$, given the complexity of protein structures and intramolecular interactions. We obviate protein size effects to an extent by only looking at mutation effects within $\gamma = \SI{15}{\angstrom}$, but there are other protein-specific factors -- such as the degree of disorder, protein shape, flexibility, and amino acid packing -- that may alter the relationship between strain and $\ddG$ for different proteins.
  Nonetheless, the strain-stability correlations shown here indicate that strain due to mutation contains considerable information about stability changes that may be leveraged in subsequent development of predictors of $\ddG$.\\

\nsi{Strain correlates with $\ddG$ almost as well as tailored $\ddG$ predictors}
  To put the strain-stability correlations in context, we compare them
  with two state-of-the-art $\ddG$ predictors, DDMut and FoldX (Appendix C)~\cite{zhona23,schna05}. FoldX predicts $\dG$ from structure using empirical energy-based potentials; it is used to calculate $\ddG$ by first a generating structure for the mutant based on a reference WT structure, which enables calculation of $\dG$ for both WT and mutant structures. DDMut uses a neural network to predict $\ddG$, using a reference structure and a mutation as input.
  
\begin{figure}
\centering
\includegraphics[width=0.95\linewidth]{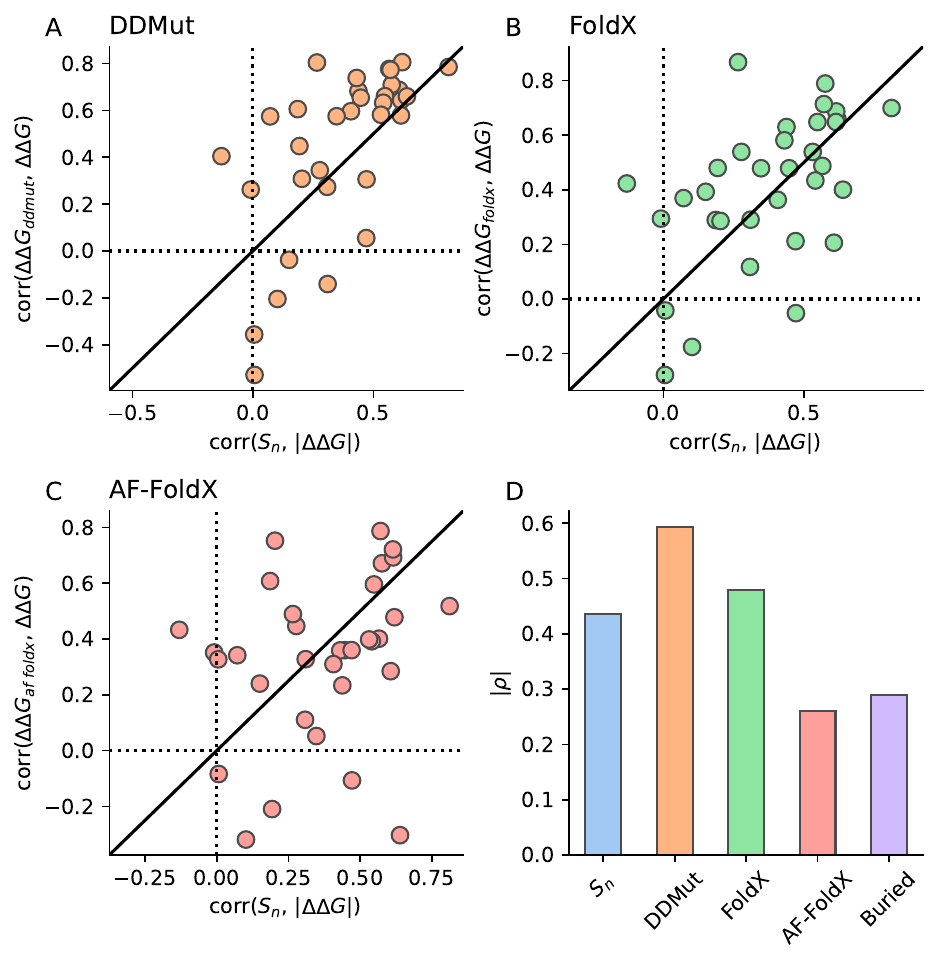}
\caption{\label{fig:fig4}
  \textbf{Comparison with state-of-the-art algorithms.}
  A-C: Correlation (Spearman's $\rho$; linear correlations are shown in SM Fig. 8) between strain $\dNn$ and magnitude of stability change $|\ddG|$, vs. correlation between $\ddG$ and predictions of DDMut (A), FoldX (B), and FoldX using 
  AF-predicted structures (C); separate points are shown for each of the 40 most-common
  proteins.
  D: Correlations (as above) for the full sample of \num{2499} unique mutants.
  Correlation between number of neighbors within \SI{10}{\angstrom} of the mutated site is shown as "Buried".
}
\end{figure}

  We find that $\dNn{-}|\ddG|$ correlations are almost as high as correlations obtained using FoldX on average, but lags behind the more recent algorithm, DDMut (\fref{fig:fig4}). We were genuinely surprised by this performance since strain is a \emph{simple} general measure of deformation and not designed specifically for stability, in contrast to FoldX and DDMut.
  However there is clearly room for improvement: simply by counting the number of neighbors at the mutation site we see a correlation of $\rho = 0.30$ (\fref{fig:fig4}D); the mean absolute error obtained from a linear fit is \SI{1.2}{kcal/mol} for both $\dNn$ and DDMut (SM~Fig.~5).
  We note that our aim here is not to use strain to predict $\ddG$, but rather to
  see whether the strain predicted by AF is informative about stability changes. But given the surprisingly high correlation, we speculate that AF-predicted structures can be leveraged to produce even better $\ddG$ predictors.

  We expected that since FoldX calculates $\dG$ for a structure,
  it would give us a more accurate estimate of $\ddG$ than strain if we use it on AF-predicted structures. Surprisingly, we find that this method (AF-FoldX) performs worse than strain in this case ($\rho = 0.26$). 
  Likewise, we find that calculating strain using the structures generated by FoldX results in a lower correlation with $\ddG$ ($\rho = 0.30$, SM~Fig.~6).
  We considered also the possibility that AF-predicted structures are not as accurate as FoldX-generated structures. However, we find that strain calculated using FoldX structures is not correlated with distance from the mutated site (SM~Fig.~7), indicating that FoldX is less accurate than AF in predicting the effect of mutations on structure.
  These results suggest that there is a promising path for generation of new energy-based methods for $\ddG$ prediction using AF-predicted structures.

\nsq{Can AF be used to predict stability changes?}
  We emphasize that our aim here is not to outright develop a $\ddG$ predictor but rather to investigate whether AF predictions are informative of stability changes. 
  We have found that a general measure of deformation, strain, correlates quite well with $\ddG$. 
  Although it was not designed to be a $\ddG$-predictor, $\dNn$ appears to be almost as good at predicting the magnitude of stability changes as state-of-the-art $\ddG$ predictors.
  Of course, this needs to be tested on a larger set of measurements, and more structures are needed for experimental validation of the relationship between strain and stability (SM Sec. 1C).
  We also note that higher correlations have been observed for these predictors on different datasets, so this analysis should be repeated on larger sets of $\ddG$ measurements~\cite{brost20,zhona23}.
  Yet, within these limitations, it seems clear that there is sufficient information in AF-predicted structures to make $\ddG$ predictions. It stands to reason that new physics-based models can be developed to achieve even better predictions, off the back of  AF-predicted structures. Alternatively, these structures could be used to reparameterize existing force fields such as FoldX. The recent explosion of high-throughput measurements~\cite{dunms21,tsuna23} will certainly lead to more machine-learning and sequence-based approaches. Nevertheless, we feel that physics-based methods are essential to offer a detailed view into the mechanistic effects of mutations on stability. The breakthrough by AF in structure prediction might offer the key to this future.

\nsi{Strain-energy relation} 
Our predictions allow us to deduce an effective strain-energy relation by modeling the protein as an elastic spring network (as detailed in SM Section 2)~\cite{tluprx17,eckrmp19}. 
Within this framework, mutation can be seen as introducing a point defect that perturbs the network \cite{dutpna18}, thereby inducing energy change $\ddG$, which includes contributions from entropy, changing topology (breaking and forming bonds), and energy release in pre-stressed frustrated bonds~\cite{ferqrb2014}.
The energy change is quadratic in the strain, with average \emph{effective} spring constants $K$ ranging between \SIrange{1}{10}{kcal/mol/\angstrom^2} (SM Fig. 9). These typical values of $K$~\cite{mcbmb22} estimate the average curvature of the high-dimensional energy landscape~\cite{eckbio21}. $K$ varies widely in and between protein families, reflecting the anisotropic geometry and heterogeneity of the landscapes.

\nsi{Advice for using AF to predict mutation effects}
  A previous study examined the same ThermoMutDB dataset, yet they concluded
  that AF cannot be used to predict stability. This is due to using 
  changes in pLDDT to measure mutation effects, which does not appear
  to be reliable for this purpose (SM~Fig.~10)~\cite{mcbpr23}.
  We recommend using strain as a more robust measure of the effect
  of mutations on structure, particularly when using averaged structures.
  We recommend using about 10 to 20 (\ie, 5 models, 2-4 repeats) structures
  to get averages, as there are diminishing returns on performance gains (SM~Fig.~11).
  Finally, we tested the strain predicted by the recently released AlphaFold3~\cite{abrna24} in three TMDB mutants by comparing with AF2 predictions (SM~Fig.~12-13), finding no major differences, except in loops and disordered regions.

  We have studied the correspondence between AlphaFold (AF) predictions of mutation effects on structure (measured using strain) and changes in stability. 
  We find that strain correlates well with $\ddG$, almost as well as state-of-the-art $\ddG$ predictors. Altogether, our findings suggest that new algorithms can be developed to extract more information from AF structures to produce accurate physics-based models of stability change upon mutation. 
  Furthermore, quantitative mapping of the energy landscape may pave the way for more realistic modeling of functional conformational dynamics of protein machines, beyond coarse-grained elastic-network models \cite{fleroy19,mcbjs24}.\\

  Code used to filter the ThermoMutDB can be found at \href{https://github.com/jomimc/AF2\_Stability\_PRL\_2024}{github.com/jomimc/AF2\_Stability\_PRL\_2024}.
  We used the \href{https://github.com/mirabdi/PDAnalysis}{PDAnalysis} python package to calculate strain.
  All AF2-predicted PDB structures are available at \cite{suppdata}, compressed using FoldComp~\cite{kimbi23}.

    This work was supported by the Institute for Basic Science, Project Code IBS-R020-D1.\\
    
\nsi{Appendix A: Stability Data}
  We select all proteins from ThermoMutDB (TMDB)~\cite{xavna21} with sequence length, $50 \leq L \leq 500$, and measurements for single mutants made within $293 \leq T \leq \SI{313}{\K}$
  and $5 \leq$ pH $\leq 8$; this amounts to \num{5078} out of \num{13337} measurements.
  One problem we found with TMDB is that the indices in mutation codes can be associated with either Protein Data Bank (PDB)~\cite{berna00}, Uniprot~\cite{unina18}, or ``unsigned'', yet we wanted to match mutations to Uniprot sequences. Hence, if indices were matched to PDB indices that differed from Uniprot indices, we used a custom script to convert the mutation codes to the correct Uniprot indices. This script only works when the mutation code index refers to the ``\_atom\_site.label\_seq\_id'' entry of the mmCIF file; we excluded many cases where the TMDB mutation indices refer to idiosyncratic indexing (\ie, not starting from one) in the ``\_atom\_site.auth\_seq\_id'' entries in the mmCIF files.
  Out of caution, if no correct matches were found for a protein we excluded all TMDB entries that were related to this protein (\ie, if they have the same value in the `PDB\_wild'' column). We excluded ``unsigned'', since these were ambiguous, and we discovered that some of these were labelled with incorrect Uniprot accession ids.
  After this procedure we are left with \num{3236} measurements. 
  When multiple $\ddG$ measurements are available we report the average $\ddG$ across all measurements. We leave out a mutant if the standard deviation of the $\ddG$ measurements is higher than \SI{1}{kT}; this occurred in only about \SI{2}{\%} of cases. 
  We examined why some of these cases had such high variance, and found occasional errors where the $\ddG$ sign differed from the reference it was taken from.
  A full list of errors and corrections can be found in SM Section 1D. The final set of $\ddG$ measurements includes \num{2499} unique mutants. We study correlations within individual wild-type (WT) proteins (and their single mutants), and across the full set of measurements.\\

\nsi{Appendix B: Structure Analysis}
  We use the ColabFold implementation of AF to predict protein structures~\cite{mirnm22}. We run all 5 models, without using templates, and using 6 recycles, and minimization using the Amber forcefield. We run 10 repeat predictions for each sequence, for each model, and create averaged structures following ~\cite{mcbpr23}. 
  We calculate effective strain, ES, which measures the deformation between mutants and WT proteins, for both averaged and non-averaged structures. ES can be described simply as the average relative change of $\CA-\CA$ distances between neighboring residues;
  we define neighbors as residues whose $\CA$ positions are within \SI{13}{\angstrom}.
  We calculate ES as~ \cite{mcbpr23},
  \begin{equation}
  \label{eq:1}
  \dNi = \left\langle \frac{|\Delta \rb_{ij}|}{|\rb_{ij}|} \right\rangle = \frac{1}{n_i}\sum_{j \in N_i} \frac{| \rb_{ij} - \rb_{ij}' |}{| \rb_{ij}|}~,
  \end{equation}
  where $\rb_{ij}$ is the distance vector between $\CA$ positions in residue $i$ and neighbor $j$ in a reference structure, $\rb'_{ij}$ is the corresponding distance vector in a target structure which has been aligned to $\rb_{ij}$, $N_i$ is the set of neighbors, and $n_i = |N_i|$ is the number of neighbors.
  We refer to ES measured at the mutated site $m$ as $\dNm$.
  We also calculate the neighborhood sum of strain $\dNn$ over residues whose distance to the mutation site $\delta_i$, is shorter than some threshold $\gamma$, 
  \begin{equation}
  \dNn = \sum_{\delta_i \le \gamma} \dNi  ~.
  \end{equation}
  We mainly use $\gamma = \SI{15}{\angstrom}$. We typically only include AF-predicted residues in strain calculations if pLDDT $> 70$,
  and treat the rest as disordered, except where otherwise noted.

\nsi{Appendix C: $\ddG$ predictors}
  We use two state-of-the-art methods to predict $\ddG$: FoldX 5~\cite{delbi19}, and DDMut~\cite{zhona23}. We use the API for the DDMut web server to predict $\ddG$. For FoldX, we use the BuildModel command to generate structures of the WT and mutant sequences and $\ddG$ predictions.
  We use five runs as recommended, and report average values of $\ddG$. For each algorithm, we provide the top-ranked (by pLDDT) AF-predicted WT structure as an input, along with a list of all mutants for that protein.

\nocite{lan86,merst99,danpn08,danpn10,robjp18,shobi90,carbi95,mcbmb22}
\bibliography{af2_stab}

\end{document}


\title{
{\smaller[0.5] Supplemental Material for} \\ ``AI-predicted protein deformation encodes energy landscape perturbation''}

  \author{John M. McBride}
    \email{jmmcbride@protonmail.com}
    \affiliation{Center for Soft and Living Matter, Institute for Basic Science, Ulsan 44919, South Korea}
  \author{Tsvi Tlusty}
    \email{tsvitlusty@gmail.com}
    \affiliation{Center for Soft and Living Matter, Institute for Basic Science, Ulsan 44919, South Korea}
    \affiliation{Departments of Physics and Chemistry, Ulsan National Institute of Science and Technology, Ulsan 44919, South Korea}

\maketitle
\tableofcontents

\section{PDB Structure Data}
\subsection{How many structures are needed to average?}
  Since we found that correlations between $\dNn$ and $\ddG$ are
  higher when using averaged AF-predicted structures, a pertinent
  practical question is, how many structures should one use?
  We looked at the correlations (both $\ddG$ and $|\ddG|$) as a
  function of the number of structures used to create average
  structures, finding that most of the improvement is seen when
  using only five structures instead of one (\fref{fig:si8}). We suggest that using
  \num{20} structures is sufficient to achieve good performance.

\subsection{pLDDT is less informative of stability than strain.}
  Since AF's predicted confidence measure, pLDDT, is a good indicator
  of disorder, we also investigated whether changes in pLDDT
  are informative of changes in stability. We compared changes in pLDDT
  at the mutated site $m$, changes in the average pLDDT (across the entire protein),
  for both single and averaged structures, against both $\ddG$ and $|\ddG|$.
  We find poor correlations between changes in pLDDT and $\ddG$ (\fref{fig:si7}), in agreement
  with \citet{pakpo23}.

\subsection{Insufficient PDB data to measure empirical strain-stability correlation.}
  We sought to repeat our analysis using experimental structures from 
  the Protein Data Bank (PDB). We only found $\sim 100$ entries which
  have both wild-type (WT) and mutant experimental structures. We removed any
  NMR structures, and only considered pairs with the same oligomeric state
  and same bound substrates. This left us with \num{36} pairs of PDB
  structures, which is too few to study statistical correlations.
  It is possible that there are sufficient $\ddG$ measurements
  across other data sets, but the limit of unique mutations
  is determined by the number of possible structure pairs that
  differ by a single amino acid substitution. In a previous study \cite{mcbpr23}
  we only found $\sim$\num{1000} of these, so there is some hope
  that this analysis can be performed if the other datasets have
  enough $\ddG$ measurements.

\subsection{Construction of a reduced ThermoMutDB set}
  Our original approach was to use the ThermoMutDB subset provided by \citet{pakpo23}, however
  we decided to work exclusively with the original ThermoMutDB data, for two
  of reasons. We found that the ``TMDB\_ID'' values do not always match with the original
  ThermoMutDB ID numbers. We also found a case where a mutant (P00644, L89V)
  was assigned an incorrect mutation position; in this case, 
  both the PDB SEQRES and Uniprot sequences have a lysine at position 89, but
  the correct mutation code should be L171V.

  We found several issues in the ThermoMutDB that led to exclusion or correction
  of entries. We found that for some entries, the reference protein for which the mutation
  was defined (and $\ddG$ was calculated) was not the WT sequence found in Uniprot;~\cite{unina18,merst99,danpn08,danpn10}
  we excluded these.
  We found duplications of entries, where measurements from one paper were referenced
  in several other papers, yet both were included;~\cite{robjp18} we excluded these entries.
  We found that one measurement (Y27A) in one source,~\cite{shobi90} and all measurements in another
  source~\cite{carbi95} were included with the incorrect sign; we corrected these.

\section{Linking spatial deformation and energy using elastic network models}

It is instructive to examine the energy landscape within a framework of effective elasticity theory, which allows us to link the scale of free energy changes with the underlying structural changes. The generic relation for a single spring is
$U = \frac{1}{2} K \Delta r^2$, where $K$ is a spring constant and $\Delta r$ is the deviation from the spring's equilibrium length. A generalization to 3D continuum material is a quadratic form for energy density,
$u = \frac{1}{2} \bf{k}_{\alpha \beta \gamma \delta }\bf{S}_{\alpha \beta } \bf{S}_{\gamma \delta}$,
where $\bf{S}$ is the $3 \times 3$ strain tensor, $\bf{k}$ is the $3\times 3 \times 3 \times 3$ elasticity tensor \cite{lan86}, and we use Einstein's summation convention.

To obtain a similar effective strain-energy relation for the protein, we model a protein as an elastic network where amino acids are connected by homogeneous springs with elastic constant $K$ between their $C_{\alpha}$ positions. Two $C_{\alpha}$ positions $i$ and $j$ are connected if they are within $\gamma$ \si{\angstrom} from each other, $|\rb_{ij}| < \gamma$. To measure the deformation, we use our previously-developed metric called the local distance difference (LDD),~\cite{mcbpr23} which is the L2-norm of the vector of changes in the $C_{\alpha}-C_{\alpha}$ distances between neighbors,
\be
    \textrm{LDD}_i = \sqrt{\sum_{j = 1}^{n_i} {\left( |\rb_{ij}| - |\rb^{'}_{ij}| \right)^2}} \, .
\ee
Here, $\rb_{ij}$ and $\rb^{'}_{ij}$ are the vectors between residue $i$ and neighbor $j$ in, respectively, a reference structure and a target structure, and the summation is over the $n_{i}$ neighbors $j$ for which $|\rb_{ij}| < \gamma$. In practice, we average $|r_{ij}|$ and $|r^{'}_{ij}|$ over many structures to get an estimate of the equilibrium distance between neighboring residues. LDD is highly correlated with effective strain~\cite{mcbpr23}, and thus is also correlated with $\ddG$ (Table \ref{tab:tab1}).
Then, the overall elastic energy of the network is simply the sum
\begin{equation}
        U = \frac{1}{4} K \sum_{i = 1}^N \sum_{j = 1}^{n_i} {\left (|\rb_{ij}| - |\rb^{'}_{ij}| \right)^2} = \frac{1}{4} K \sum_{i = 1}^N {\textrm{LDD}_i^2}\, ,
        \label{eq:U}
\end{equation}
where the additional $\frac{1}{2}$ factor accounts for counting twice each $i-j$ bond.

To get a rough estimate for the typical values of \textit{effective} spring constant $K$ that describes the energy landscape of the protein, we first compute the LDD by comparing wild-type and mutated AF-predicted structures, exactly in the manner in which the effective strain is computed (see App. B in manuscript). Next, we equate the change in stability $\ddG$ to the elastic energy $U$ in Eq. (\ref{eq:U}), from which we obtain an approximation for $K$,
  \begin{equation}
          K =   \frac{4 \,\ddG}{\sum \limits_{i} \textrm{LDD}_i^2}~.
          \label{eq:K}
  \end{equation}
  
  It is essential to note that $\ddG$ includes significant contributions from changes in the entropy and the topology of the bond network (bonds may break and reform), which may sometimes decrease the overall energy. Moreover, many bonds in the equilibrium conformation of the protein are ``frustrated''~\cite{ferqrb2014}, \ie~shorter or longer than their equilibrium length. Hence, deformation can release some of the elastic energy in these pre-stressed bonds. For all these reasons, the estimated spring constant $K$ from \eref{eq:K} is only an effective parameter, which serves as a ballpark figure for the actual strength of elastic interactions between amino acids.
 
  Using this method, we estimate $K$ for each of the \num{46} protein families for which we have \num{10} or more $\ddG$ values. We calculate the sum of squared $\textrm{LDD}_i$ for each protein in a family, fit a straight line between this and $\ddG$ using \eref{eq:U} with an intercept at zero, and get $K$ from the gradient.
  As expected, the value of $K$ depends strongly on the topology of the assumed bond network, in which the number of springs is determined by the neighborhood radius $\gamma$. Choosing $\gamma \sim  \SIrange{6}{7}{\angstrom}$ corresponds to bonds between amino acids that are likely to be directly in physical contact. At this range, we find an average spring strength of $K \sim \SI{10}{kcal/mol/\angstrom^2}$ (\fref{fig:si6c}). Choosing Larger neighborhood radii $\gamma \sim  \SIrange{10}{11}{\angstrom}$, as commonly assumed in elastic network models ~\cite{mcbmb22}, yields lower average values of $K \sim \SI{1}{kcal/mol/\angstrom^2}$. The effective $K$ values vary broadly, with a standard error of about one order of magnitude, reflecting the diverse, intricate geometry of protein energy landscapes. Even in a single protein family, the highly anisotropic landscape exhibits a wide distribution of spring constants. Certain directions in the space of conformations are very flat as the corresponding deformation hardly changes $\ddG$, while other directions are very steep with large $\ddG$ changes  ~\cite{mcbmb22}.

  \begin{figure*}[t!]  \centering
  \includegraphics[width=0.99\linewidth,]{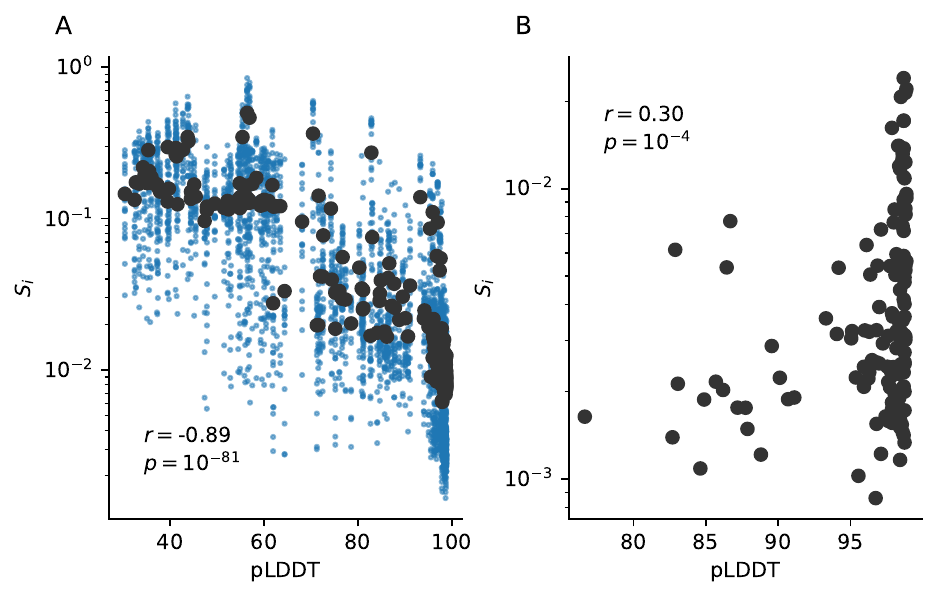}
  \caption{\label{fig:si1}
  \textbf{AF predictions are more variable in regions with low pLDDT.}
  Strain per residue $\dNi$ vs. pLDDT per residue for:
  (A) Strain calculated between repeat AF predictions of the same protein sequence;
  blue dots are shown for different pairs of single structures; black circles
  indicate means across multiple pairs.
  (B) Strain calculated for pairs of averaged structures, whose sequences differ
  by a single mutation; residues with pLDDT $<70$ are excluded.
  Pearson's correlation coefficient and p value are shown.
}
  \end{figure*}

  \begin{figure*}[t!]  \centering
  \includegraphics[width=0.90\linewidth,]{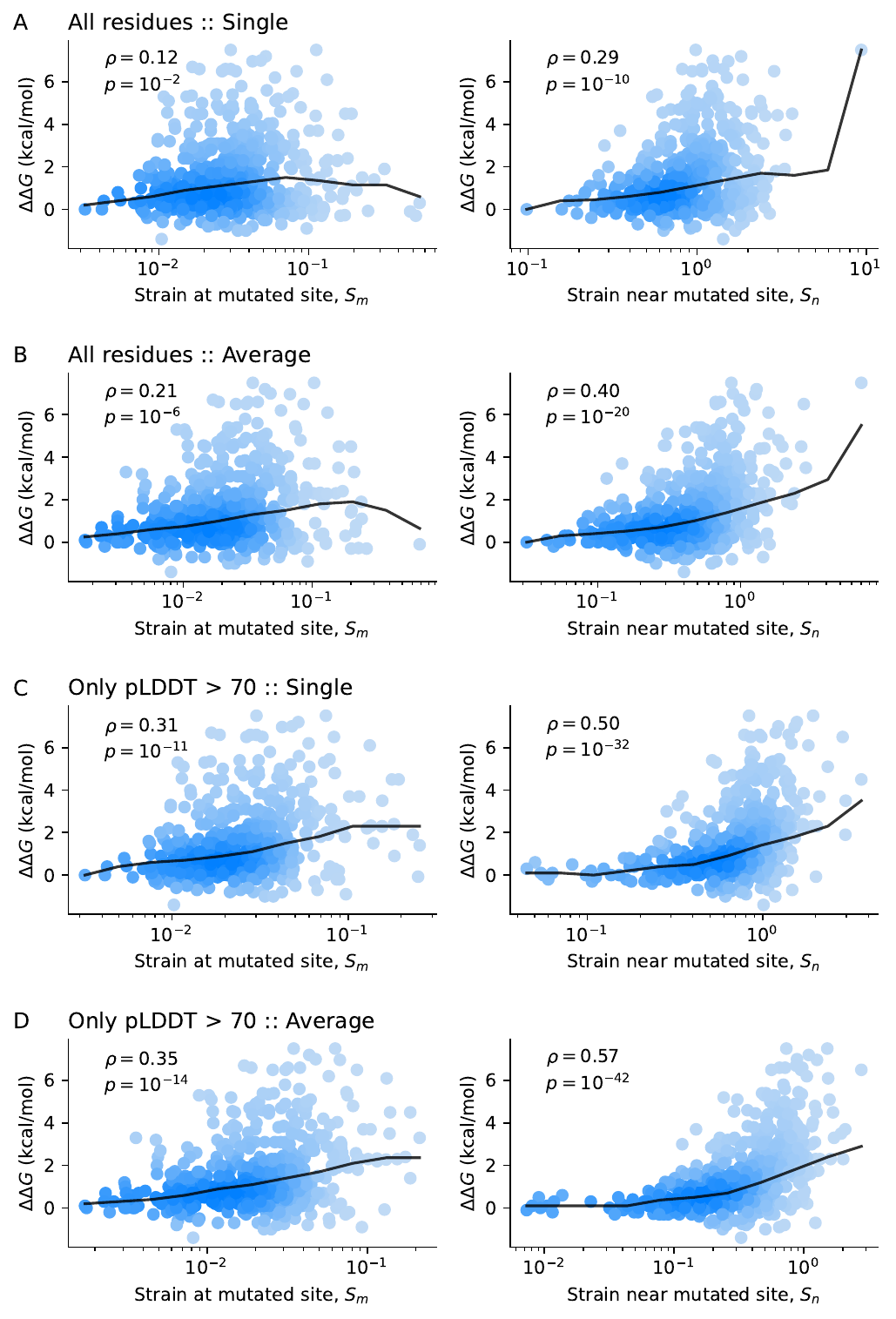}
  \caption{\label{fig:si2}
  \textbf{Improving precision of mutation effect prediction.}
  Strain at the mutated site $m$, $\dNm$ (left), and near the
  mutated site, $\dNn$ (right), for four different methods of calculating
  strain: (A) All residues and pairs of single structures.
  (B) All residues and pairs of averaged structures.
  (C) Only residues with pLDDT $>70$ and pairs of single structures.
  (D) Only residues with pLDDT $>70$ and pairs of averaged structures.
  Spearman's $\rho$ and p value are shown.
}
  \end{figure*}

  \begin{figure*}[t!]  \centering
  \includegraphics[width=0.99\linewidth,]{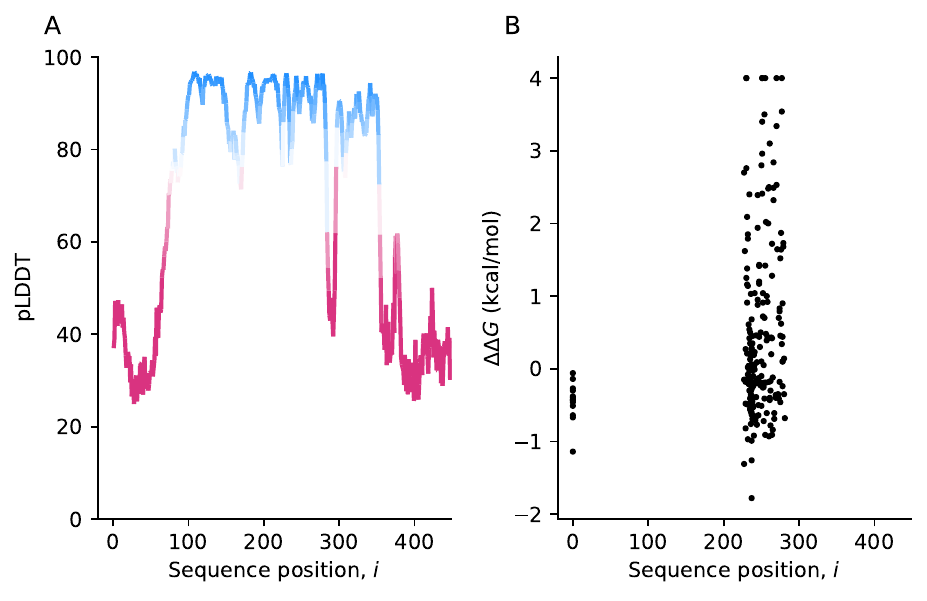}
  \caption{\label{fig:si3}
  \textbf{Biased sampling of IgG-binding protein G (IgGb).}
  A: pLDDT vs sequence position $i$ for IgGb.
  B: Distribution of mutations in the ThermoMutDB for IgGb along
  the sequence, and the corresponding $\ddG$ values.
}
  \end{figure*}

  \begin{figure*}[t!]  \centering
  \includegraphics[width=0.99\linewidth,]{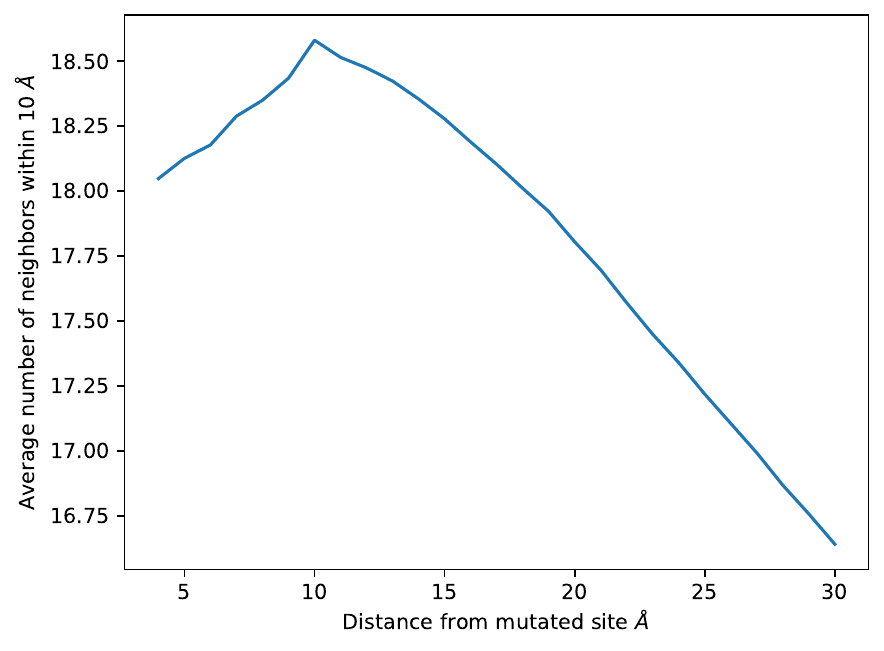}
  \caption{\label{fig:si4}
  \textbf{The most buried residues tend to be within \SI{15}{\angstrom} of other residues.}
  For each protein we measure the number of neighbors per residue. Neighbors are defined as the set of residues whose
  $\CA$ distances are within \SI{10}{\angstrom}. Plot shows average number of neighbors of residues
  as a function of the distance to the mutated site.
  Average is calculated for all unique proteins in our reduced ThermoMutDB sample.
}
  \end{figure*}

  \begin{figure*}[t!]  \centering
  \includegraphics[width=0.99\linewidth,]{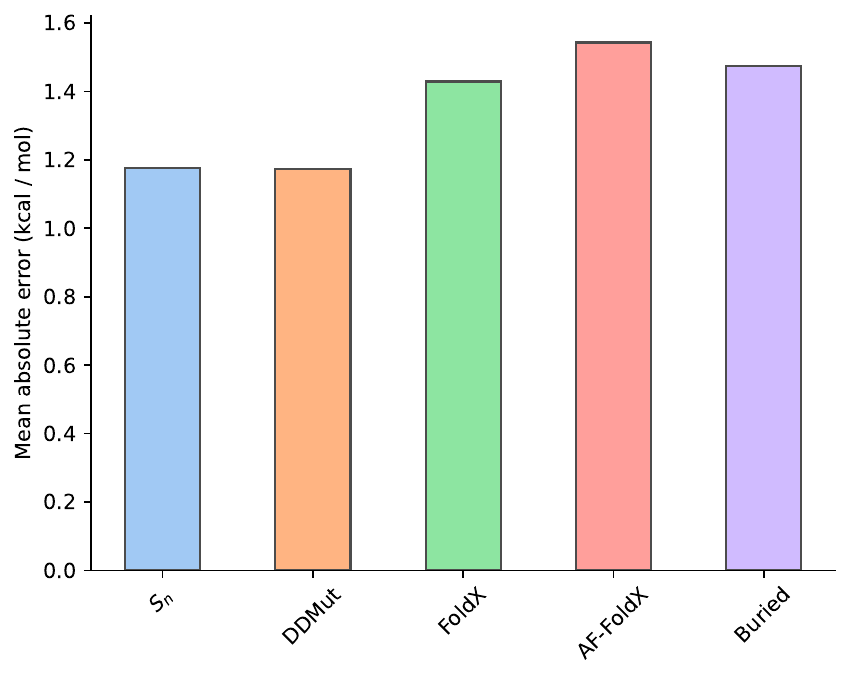}
  \caption{\label{fig:si4b}
  Mean absolute error of linear fits to $\ddG$ using various methods.
}
  \end{figure*}

  \begin{figure*}[t!]  \centering
  \includegraphics[width=0.99\linewidth,]{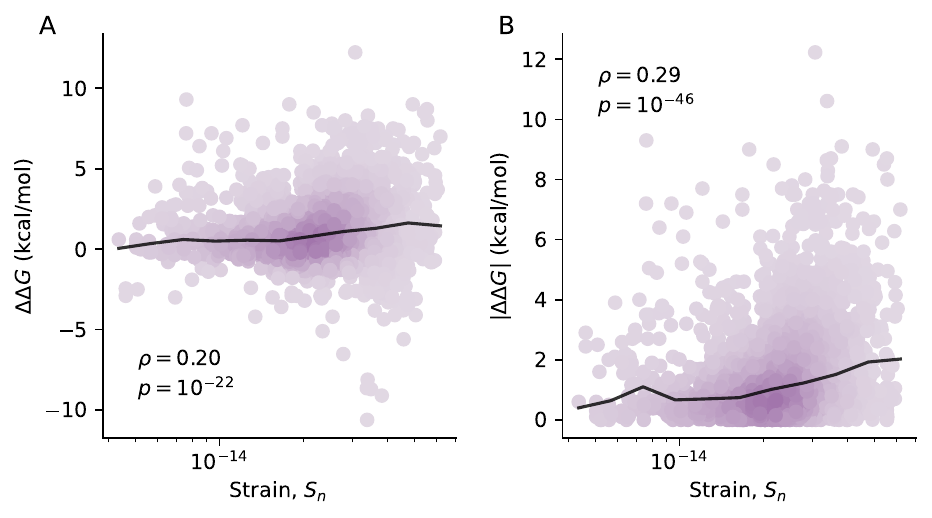}
  \caption{\label{fig:si5}
  \textbf{$\dNn$--$\ddG$ correlations calculated using FoldX structures.}
  Strain near mutated site $\dNn$ vs. $\ddG$ (A) and magnitude of stability
  change $|\ddG|$ for all mutations in our reduced ThermoMutDB sample
  of \num{2499} unique mutants. Spearman's $\rho$ and p value are shown.
}
  \end{figure*}

  \begin{figure*}[t!]  \centering
  \includegraphics[width=0.99\linewidth,]{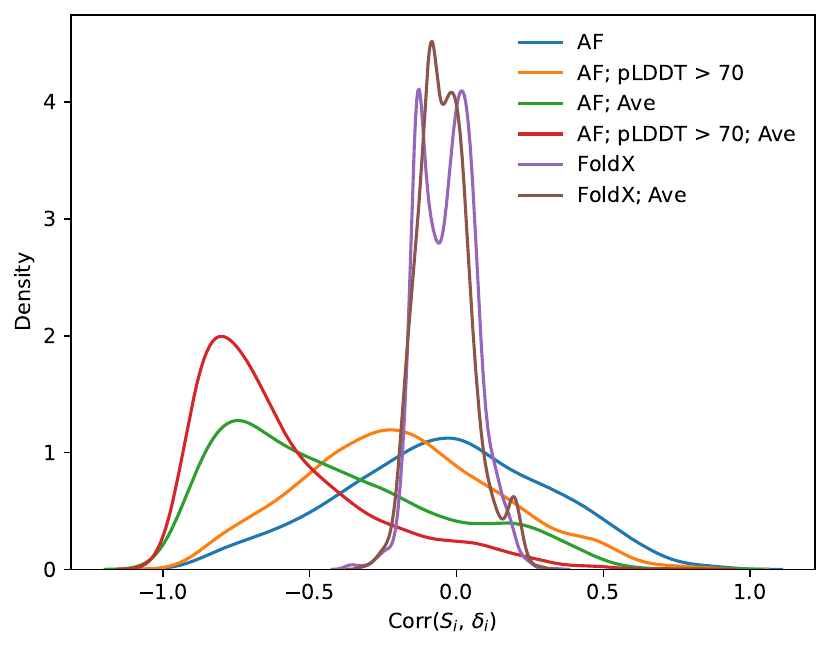}
  \caption{\label{fig:si6}
  \textbf{Correlation between distance from mutated site $\delta_i$ and $\dNi$}
  for all proteins in our reduced ThermoMutDB sample.
  Different distributions are shown for shear calculated using:
  AF-predicted structures, all residues and pairs of single structures (AF).
  AF-predicted structures, only residues with pLDDT $>70$ and pairs of single structures (AF; pLDDT $> 70$).
  AF-predicted structures, all residues and pairs of averaged structures (AF; Ave).
  AF-predicted structures, only residues with pLDDT $>70$ and pairs of averaged structures (AF; pLDDT $> 70$; Ave).
  FoldX-predicted structures, all residues and pairs of single structures (FoldX).
  FoldX-predicted structures, all residues and pairs of averaged structures (FoldX; Ave).
}
  \end{figure*}

  \begin{figure*}[t!]  \centering
  \includegraphics[width=0.99\linewidth,]{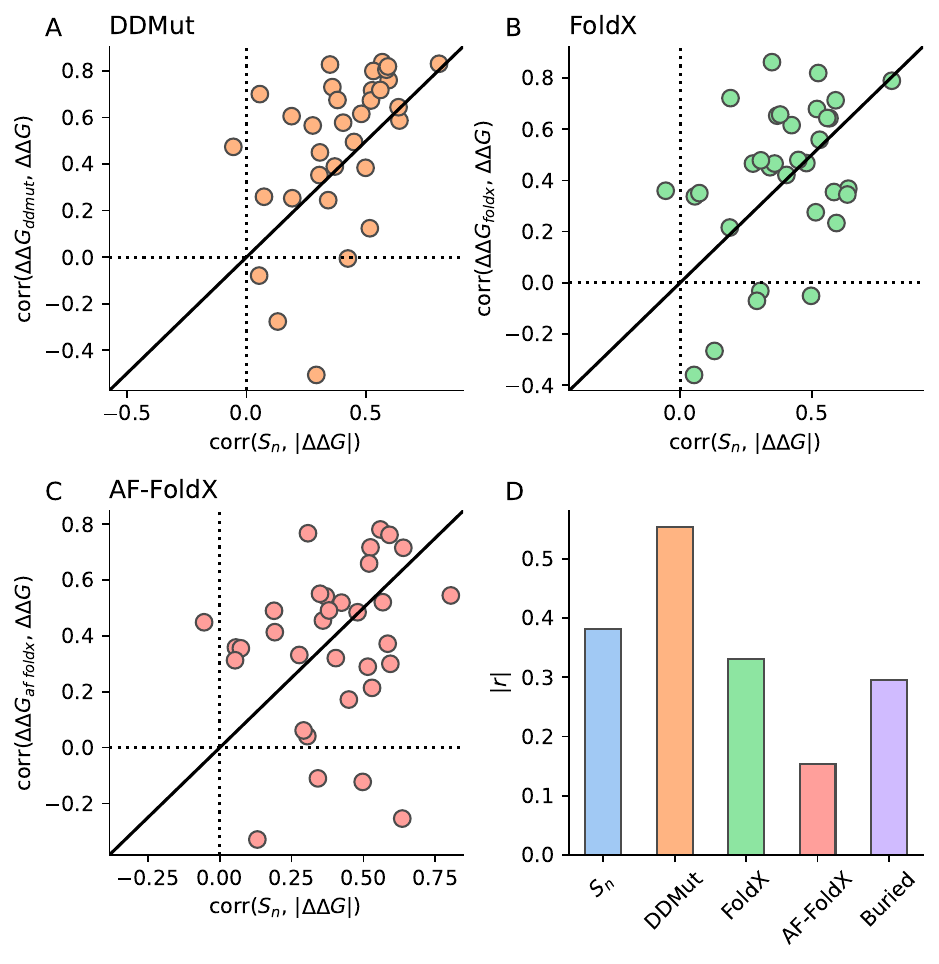}
  \caption{\label{fig:si6b}
  \textbf{Comparison with state-of-the-art algorithms.}
  A-C: Correlation (Pearson's $r$) between strain $\dNn$ and magnitude of stability change $|\ddG|$, vs. correlation between $\ddG$ and predictions of DDMut (A), FoldX (B), and FoldX using 
  AF-predicted structures (C); separate points are shown for each of the 40 most-common
  proteins.
  D: Correlations (as above) for the full sample of \num{2499} unique mutants.
  Correlation between number of neighbors within \SI{10}{\angstrom} of the mutated site is shown as "Buried".
}
  \end{figure*}

  \begin{figure*}[t!]  \centering
  \includegraphics[width=0.99\linewidth,]{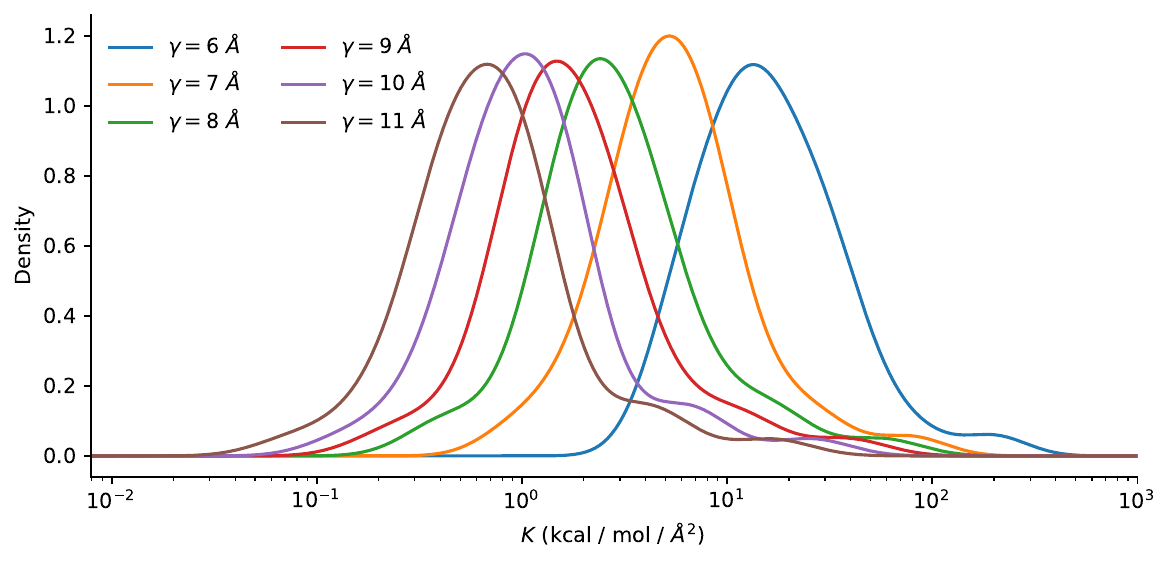}
  \caption{\label{fig:si6c}
  \textbf{Computation of effective spring constant from stability changes.}
  Distribution of homogeneous spring constant $K$ inferred from empirical $\ddG$ and predicted deformation using an elastic network model (\eref{eq:K}). Different values of $K$ are obtained for different elastic network topologies, defined by the neighborhood cutoff radius $\gamma$ \si{\angstrom}.
}
  \end{figure*}

  \begin{figure*}[t!]  \centering
  \includegraphics[width=0.99\linewidth,]{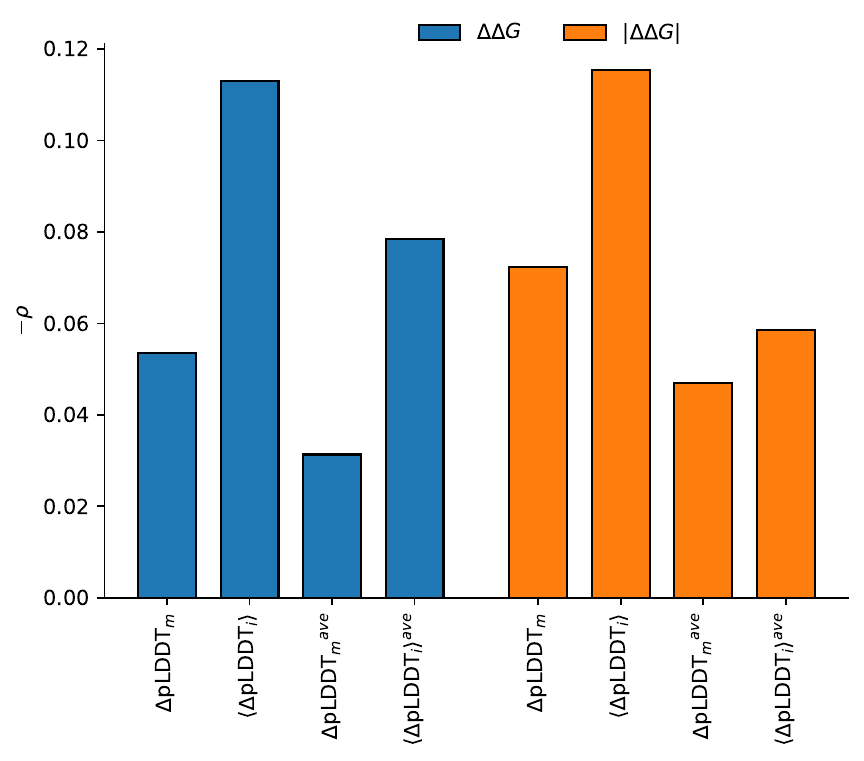}
  \caption{\label{fig:si7}
  \textbf{$\Delta$pLDDT is a poor predictor of $\ddG$.}
  Spearman's correlation between differences in pLDDT and stability.
  Differences in pLDDT are measured in four ways:
  difference in pLDDT at the mutated site $\dP_m$;
  average difference in pLDDT across the whole protein $\langle \dP_i \rangle$;
  difference in pLDDT at the mutated site, averaged over 50 pLDDT predictions $\dP_m^{ave}$;
  average difference in pLDDT across the whole protein,
  averaged over 50 pLDDT predictions $\langle \dP_i \rangle^{ave}$;
}
  \end{figure*}

  \begin{figure*}[t!]  \centering
  \includegraphics[width=0.99\linewidth,]{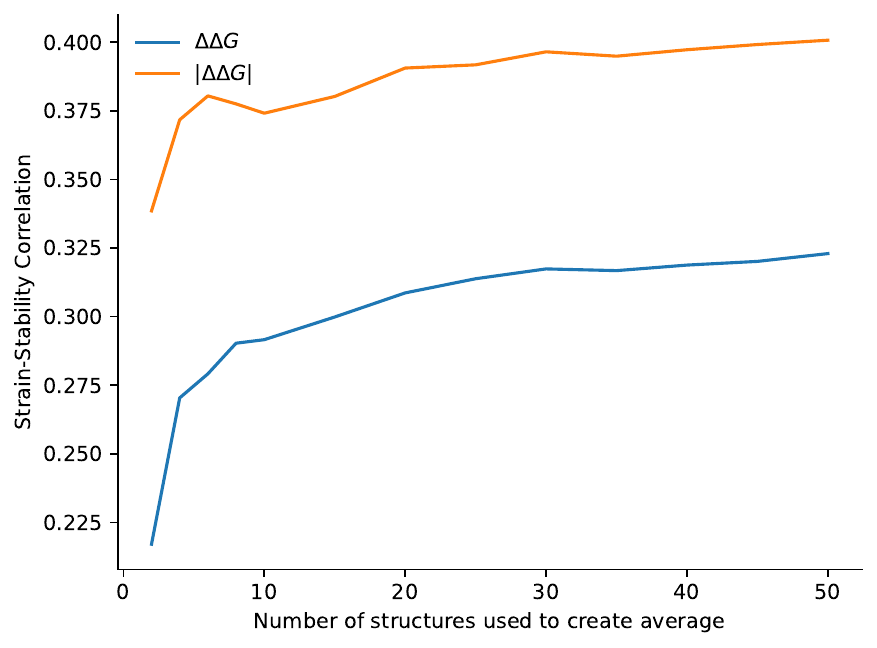}
  \caption{\label{fig:si8}
  \textbf{Effect of the number of structures used to create averages.}
  Strain-stability correlation for all mutations in our reduced ThermoMutDB sample
  of \num{2499} unique mutants, as a function of the number of structures used
  to create average structures.
}
  \end{figure*}

  \begin{figure*}[t!]  \centering
  \includegraphics[width=0.99\linewidth,]{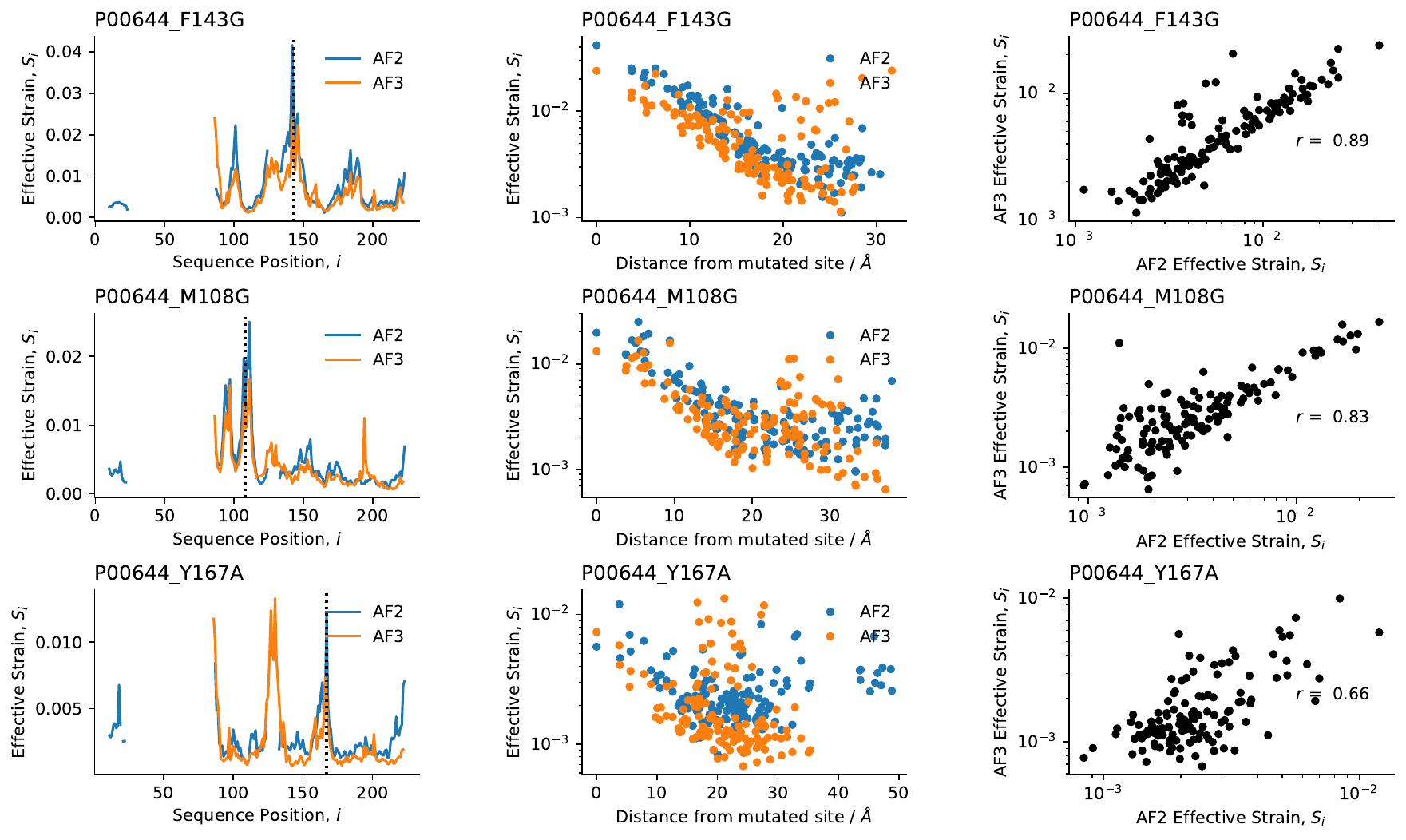}
  \caption{\label{fig:si12}
  \textbf{Comparison of AlphaFold2 (AF2) and AlphaFold3 (AF3) strain predictions.}
  Strain is calculated by AF2 and AF3~\cite{abrna24} between Themonuclease WT (P00644) and three mutants: F143G (top), M108G (middle) and Y167A (bottom).
  Left: Strain (AF2 and AF3) as a function of sequence position.
  Center: distance from the mutated site (center).
  Right: AF3-predicted strain against AF2-predicted strain (Pearson's $r$ is shown).
  The strain values are generally similar, except for some flexible regions where AF2 and AF3 give different pLDDT values (\eg, the loop around residue 128, see SI \fref{fig:si13}).
}
  \end{figure*}

  \begin{figure*}[t!]  \centering
  \includegraphics[width=0.99\linewidth,]{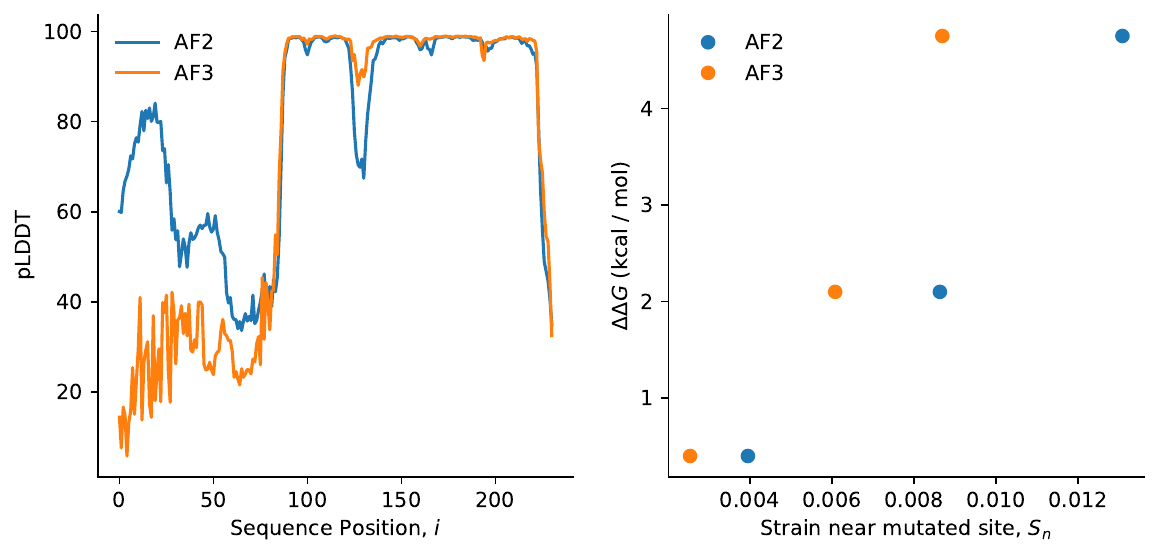}
  \caption{\label{fig:si13}
  \textbf{Comparison of AF2 and AF3 pLDDT and mutation effect on stability.}
  Left: pLDDT (AF2 and AF3) vs. sequence position. 
  The discrimination between ordered (pLDDT > 70) and disordered (pLDDT < 70) regions is generally similar, except for the loop around residue 128 (for which AF2 predicts pLDDT $\sim$ 70 and AF3 pLDDT $\sim$ 90). 
   Right: Change in stability upon mutation, $\ddG$, vs. strain near the mutated site, $S_n$, for three Thermonuclease mutants (F143G, M108G, Y167A).
  }
  \end{figure*}

\clearpage

\begin{table}
\centering
\caption{\label{tab:tab1}Comparison of linear and rank-order correlations between $\ddG$ and either Effective Strain or LDD (summation over residues near the mutated site), for main text Figures 1-3.}
\begin{tblr}{
  hline{2} = {-}{},
}
Figure      & Correlation & Strain & LDD  \\
Fig1F       & Linear      & 0.35   & 0.4  \\
Fig1F       & Rank Order  & 0.35   & 0.41 \\
Fig1G       & Linear      & 0.52   & 0.51 \\
Fig1G       & Rank Order  & 0.57   & 0.55 \\
Fig2A (i)   & Linear      & 0.44   & 0.42 \\
Fig2A (i)   & Rank Order  & 0.37   & 0.36 \\
Fig2A (ii)  & Linear      & 0.6    & 0.59 \\
Fig2A (ii)  & Rank Order  & 0.61   & 0.59 \\
Fig2A (iii) & Linear      & 0.48   & 0.45 \\
Fig2A (iii) & Rank Order  & 0.52   & 0.47 \\
Fig2A (iv)  & Linear      & 0.5    & 0.49 \\
Fig2A (iv)  & Rank Order  & 0.57   & 0.54 \\
Fig3A       & Linear      & 0.31   & 0.29 \\
Fig3A       & Rank Order  & 0.36   & 0.34 \\
Fig3B       & Linear      & 0.37   & 0.35 \\
Fig3B       & Rank Order  & 0.41   & 0.4  
\end{tblr}
\end{table}

\clearpage
\bibliography{af2_stab}